\documentclass[12pt]{article}
\usepackage[english,activeacute]{babel}
\usepackage{natbib}
\usepackage{comment}
\usepackage{float}
\usepackage[hidelinks]{hyperref}
\usepackage{mathrsfs}
\usepackage{enumitem}
\usepackage[font={small,it}]{caption}
\usepackage{amsmath,amsfonts,amsthm,amssymb}
\usepackage{bm,rotating,multirow,dsfont,graphicx}
\usepackage[usenames,dvipsnames]{color}
\usepackage{url}
\usepackage{multicol}
\usepackage{multirow}
\usepackage[T1]{fontenc}
\usepackage{flafter}
\usepackage{appendix}
\usepackage{subfigure}
\usepackage{xcolor}
\usepackage{soul}
\usepackage{threeparttable}
\usepackage{booktabs}
\makeatletter
\def\hlinewd#1{%
	\noalign{\ifnum0=`}\fi\hrule \@height #1 %
	\futurelet\reserved@a\@xhline}
\makeatother
\newcommand{\be}{\begin{eqnarray*}}
\newcommand{\ee}{\end{eqnarray*}}
\newcommand{\bet}{\begin{eqnarray}}
\newcommand{\eet}{\end{eqnarray}}
\def\spacingset#1{\renewcommand{\baselinestretch}{#1}\small\normalsize}\spacingset{1}
\def\@roman#1{\romannumeral #1}
\graphicspath{{./plots/}}

\begin{document}

\title{An unified approach to link prediction in \\ collaboration networks}

\author{
    Juan Sosa$^{1}$\footnote{Email: jcsosam@unal.edu.co.}$\qquad$
    Diego Martínez$^{1}$\footnote{Email: damartinezsi@unal.edu.co.}$\qquad$
    Nicolás Guerrero$^{1}$\footnote{Email: niguerreroc@unal.edu.co.}
}

\date{
    $^{1}$Universidad Nacional de Colombia, Bogotá, Colombia
}

\maketitle


\begin{abstract} 
\noindent
This article investigates and compares three approaches to link prediction in colaboration networks, namely, an ERGM (Exponential Random Graph Model; \citealt{robins2007introduction}), a GCN (Graph Convolutional Network; \citealt{kipf2017semi}), and a Word2Vec+MLP model (Word2Vec model combined with a multilayer neural network; \citealt{mikolov2013efficient} and \citealt{Goodfellow}). The ERGM, grounded in statistical methods, is employed to capture general structural patterns within the network, while the GCN and Word2Vec+MLP models leverage deep learning techniques to learn adaptive structural representations of nodes and their relationships. The predictive performance of the models is assessed through extensive simulation exercises using cross-validation, with metrics based on the receiver operating characteristic curve. The results clearly show the superiority of machine learning approaches in link prediction, particularly in large networks, where traditional models such as ERGM exhibit limitations in scalability and the ability to capture inherent complexities. These findings highlight the potential benefits of integrating statistical modeling techniques with deep learning methods to analyze complex networks, providing a more robust and effective framework for future research in this field.
\end{abstract}

\noindent
{\it Collaboration networks, Exponential Random Graph Model, Graph Convolutional Network, Word2Vec, Social Networks Analysis.}

\spacingset{1.1} 

\section{Introduction}\label{intro}

Social networks have captured the attention of researchers since the last century. Studying networks is essential, as we live in a connected world where understanding connections provides deeper insights into numerous phenomena \citep{kolaczyk_statistical_2020}. A specific area of interest within this field is collaboration networks. While some authors focus on studying these networks as relationships formed between actors who interact locally, resulting in biased and random components \citep{skvoretz_biased_1990}, we consider a collaboration network as a fully structured complex system in which two scientists are connected if they have coauthored an article. 
This definition is reasonable, as most people who have written an article together have shared ideas over a certain period \citep{newman_structure_nodate}. Some researchers have analyzed collaboration graphs for scientists across various fields, using data from sources such as MEDLINE, the Los Alamos e-Print Archive, SPIRES, and NCSTRL \citep{newman_structure_nodate}, focusing primarily on descriptive studies and certain clustering processes.

In the context of collaboration networks, a primary objective is to predict new links within the network. Exponential Random Graph Models (ERGM; e.g., \citealt{snijders2002markov}, \citealt{robins2007introduction} and \citealt{lusher2012exponential}) are commonly used for this purpose, although some applications have focused on economic collaboration networks, given that economic development has been driven by competitiveness \citep{lee_interorganizational_2012}. The use of these models is justified by their recognition as the most powerful, flexible, and widely applied approach for constructing and testing statistical network models \citep{luke_users_2015}. However, machine learning models have advanced significantly, and their applications to networks have been extensively explored in recent years. This trend has led to two distinct perspectives: Models that prioritize the network’s structure and models that represent networks as a set of vectors (e.g., \citealt{hamilton2017representation} and \citealt{zhang2020deep}).

On the one hand, Graph Convolutional Network (GCN) models stand out (e.g., \citealt{kipf2017semi}, \citealt{hamilton2017inductive}, and \citealt{wu2021comprehensive}). This approach extends the convolution operation to graphs, enabling the network to learn node representations by considering both node features and graph structure \citep{yao_graph_2018}. Through convolutional layers, nodes aggregate information from their neighbors, allowing the model to capture local relationships and structural patterns within the graph. This capability makes GCNs particularly useful for tasks such as node classification \citep{kipf2017semi}, link prediction, and clustering \citep{chiang_cluster-gcn_2019}, as they can infer and generalize the complex relationships inherent in relational data.

On the other hand, embedding algorithms are essential. The primary goal of graph embedding methods is to encode nodes within a latent vector space, effectively capturing each node's properties in a lower-dimensional vector form \citep{xu_understanding_2021}. This representation aims to preserve the overall structure of the graph, positioning related nodes with similar characteristics as ``close'' vectors. This approach is essentially similar to Hoff's latent space models (e.g., \citealt{hoff2002latent}, \citealt{hoff2007modeling}, and \citealt{sosa2021review}). Typically, continuous-space language models are employed for this purpose, which focus on representing words as vectors \citep{mikolov_linguistic_nodate}. Given the analogy between graphs and word sequences, these models can be adapted to network contexts by systematically traversing each edge and vertex in the graph \citep{skiena_algorithm_2008} and by categorizing the importance of nodes within the network \citep{skiena_data_2017}. This approach results in a set of vectors, enabling the application of all common properties and operations of a traditional vector space.

Despite the popularity of these modeling approaches, to the best of our knowledge, a comprehensive comparison of these tools for link prediction remains unavailable. Therefore, this study systematically compares the predictive performance of Exponential Random Graph Models (ERGMs), Graph Neural Networks (GNNs), and embedding models across five academic collaboration networks. The objective is to determine which approach offers the highest accuracy and computational efficiency and to provide clear recommendations for their application across various contexts. Although the analysis focuses on the specific case of the Astro-Ph (Astrophysics) network, as it is the largest and densest among the five networks, the findings are shown to be consistent across the other cases as well.

This study is crucial, as understanding the specific strengths and limitations of each modeling method is imperative for advancing network analysis and developing more effective techniques for predicting relationships in collaboration networks and other relational datasets with similar structures.
The scope of this project encompasses a detailed review and careful implementation of these models (which is not a straightforward task!) followed by a comparison using specialized predictive metrics, and the generation of recommendations based on the results, with a focus on academic collaboration networks.

The remainder of the article is organized as follows. Section 2 provides a detailed overview of the fundamental aspects of the models under study. Section 3 offers a comprehensive comparison of the models using five well-known academic collaboration networks. Finally, Section 4 presents the findings along with several recommendations for future research.

\section{Modeling}

In this section, we present the most relevant details regarding the models to be evaluated for predictive purposes, namely, Exponential Random Graph Models (ERGMs), Graph Neural Networks (GNNs), and embedding models, including their corresponding theoretical and computational details.

Here’s the translation and revision of your section into academic English, with enhanced clarity and structure:

\subsection{Exponential Random Graph Model}

The Exponential Random Graph Model (ERGM) is a statistical framework used to represent and analyze complex networks, allowing for the capture of dependencies between links and the characteristics of the nodes that compose the network. 
Unlike simpler models, such as the Erdős-Rényi model \citep{erdos1960evolution} and the generalized random graph model \citep{newman2001random}, ERGMs allow for the inclusion of structural dependencies, such as tendencies toward triangle formation or the preference for certain nodes to be more highly connected \citep{lusher_definition_2013}. 
Fundamentally, ERGMs model the probability of observing specific link formations in a network based on a set of parameters associated with both nodal and network statistics. One clear advantage of this approach is the interpretability of its parameters, which aids in understanding the underlying factors influencing link formation in a particular network. However, in large networks, the computational cost of estimating ERGM parameters can be substantial, which complicates the process and often leads to the use of alternative algorithms \citep{handcook_computer_intensive_2008}.

An ERGM is formally defined as
\begin{equation*}
p\left(\mathbf{y}\mid \boldsymbol{\theta}\right) = 
\frac{1}{\kappa} \exp{ \{ \boldsymbol{\theta}^\textsf{T} \mathbf{g(y)} \} }\,,
\end{equation*}
where $\mathbf{y} = [y_{i,j}]$ represents the realization of a random adjacency matrix \(\mathbf{Y}\), \(\mathbf{g(y)}\) is a \(K\)-dimensional vector of network statistics (endogenous variables) and vertex characteristics (exogenous variables), \(\boldsymbol{\theta}\) is a \(K\)-dimensional vector of unknown parameters, and \(\kappa \equiv \kappa(\boldsymbol{\theta})\) is the normalizing constant ensuring that $p\left(\cdot\mid \boldsymbol{\theta}\right)$ is a proper probability distribution.

ERGM functionality involves specifying relevant network statistics and estimating the corresponding parameters, which indicate the relative significance of each statistic within the network's relational structure. 
Due to the complexity of normalizing probabilities across all possible networks, parameter estimation is typically performed using Markov Chain Monte Carlo (MCMC) methods (see \citealt{gamerman2006markov} and \citealt{robins2007introduction}), which can be slow depending on the network size.
ERGMs are particularly valuable when networks are assumed to form not only through random connections but also through significant structural patterns that can be captured and analyzed using both endogenous and exogenous variables.

\subsection{Graph Convolutional Network}

Graph Convolutional Networks (GCNs) are neural networks \citep{Goodfellow} specifically designed to process graph-structured data directly, without transforming the graph. Unlike traditional convolutional networks, which operate on grid-based data like images, GCNs extend the convolution operation to graphs, capturing relationships between nodes and edges. This enables effective learning for tasks like node classification and link prediction in complex graph-structured data.

\begin{figure}[htb]
    \centering
    \includegraphics[width = 0.9\textwidth]{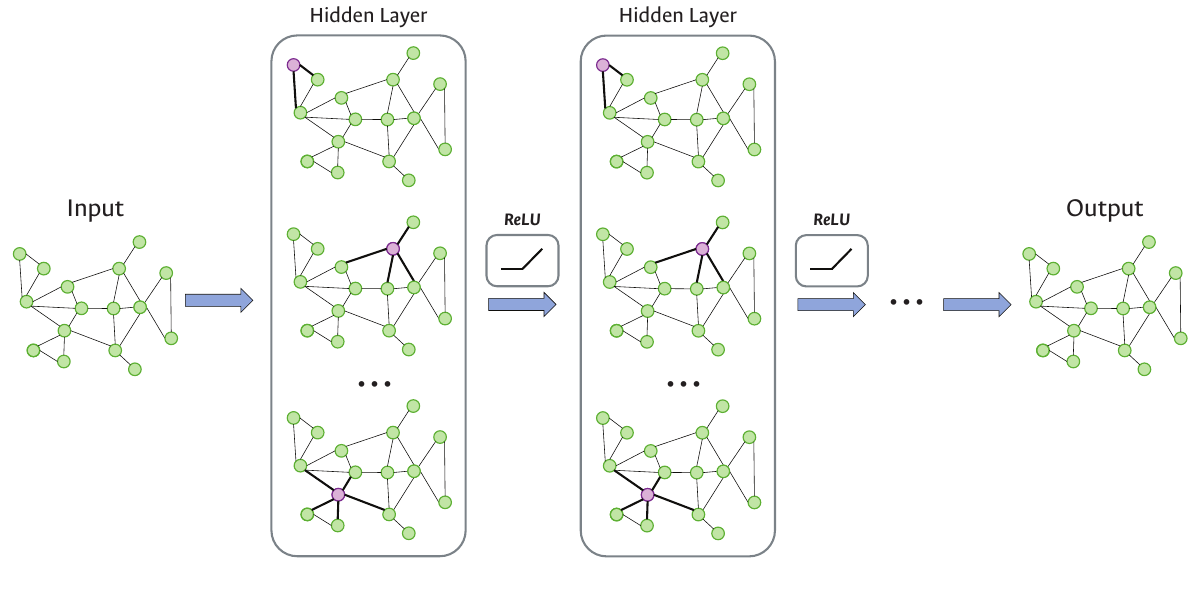}
    \caption{GCN Model: Adapted from \cite{kipf2017semi} and \cite{xu_understanding_2021}.}
    \label{GNC}
\end{figure}

Graph convolution allows each node to aggregate and process information from not only its own features but also from those of its neighboring nodes, as illustrated in Figure \ref{GNC}. This capability is essential for capturing local relationships and structural patterns in the graph. In this context, the convolution operation is defined by \cite{kipf2017semi} as:
\[
\mathbf{H}^{(l+1)} = \sigma \left( \tilde{\mathbf{D}}^{-1/2}\, \tilde{\mathbf{A}}\, \tilde{\mathbf{D}}^{-1/2}\, \mathbf{H}^{(l)}\, \mathbf{W}^{(l)} \right)\,,
\]
where \( \mathbf{H}^{(l)} \) is the feature matrix in layer \( l \), \( \tilde{\mathbf{A}} = \tilde{\mathbf{A}} + \mathbf{I} \) is the adjacency matrix with added self-connections (\( \mathbf{I} \) is the identity matrix), \( \tilde{\mathbf{D}} \) is the diagonal degree matrix of \( \tilde{\mathbf{A}} \), \(\mathbf{W}^{(l)} \) is the weight matrix in layer \( l \), and \( \sigma \) is an activation function, such as ReLU, which provides non-linearity, helping the model capture complex relationships across nodes.

Graph-level outputs can be modeled by introducing pooling operations that aggregate node information across the graph structure \cite{duvenaud_convolutional_2015}. By calculating these representations as weighted combinations of the features of neighboring nodes, pooling operations help capture information from multiple scales within the graph, allowing the model to abstract details progressively. This enables the network to learn complex structures in a hierarchical and layered manner \cite{kipf2017semi}, supporting tasks that require an understanding of the graph as a whole, such as graph classification or property prediction.

\subsection{Word2Vec}

Word2Vec is a deep learning model widely employed in natural language processing (NLP; e.g., \citealt{Amarasinghe2024}) to learn continuous vector representations of words within a high-dimensional space. As its name suggests (``Word to Vector''), the model uses neural networks to transform \textit{words} into \textit{vectors}, positioning words with similar contexts close to each other in the vector space \citep{mikolov_linguistic_nodate}. One commonly used method is the skip-gram approach \citep{Silge2017}, which aims to predict the context surrounding a given word \citep{xu_understanding_2021}. The skip-gram model's loss function is defined as follows:
\begin{equation*}
\mathcal{L}_{\text{Skip-gram}} = -\displaystyle\sum_{t=1}^{T} \displaystyle\sum_{-c \leq j \leq c} \log p(w_{t+j} \mid w_t)\,,
\end{equation*}
where \( w_t \) is the target word at time \( t \), \( c \) is the context window size, and \( p(w_O \mid w_I) \) is the probability of observing the word \( w_O \) given the context word \( w_I \), modeled through a neural network.

\begin{figure}[htb]
    \centering
    \includegraphics[width = 0.9\textwidth]{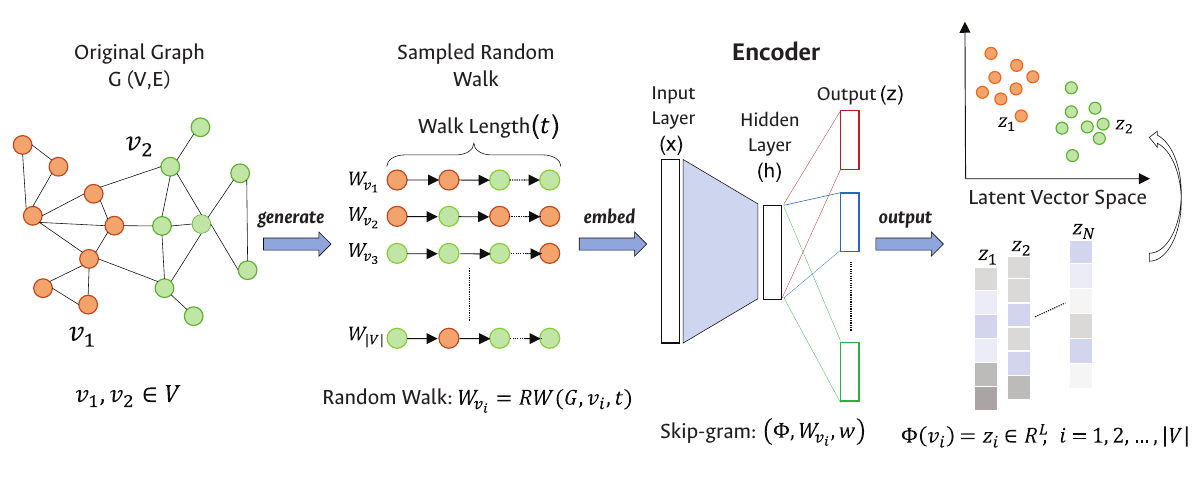}
    \caption{Embedding model: Adapted from \cite{xu_understanding_2021}.}
    \label{Word2vec}
\end{figure}

This model has been widely applied in natural language processing (NLP) but it is also adaptable to social network analysis due to its ability to transform high-dimensional, non-Euclidean spaces into lower-dimensional vector spaces \citep{xu_understanding_2021}. In this context, \textit{sentences} are represented by random walks on the target graph, following existing edges, as illustrated in Figure \ref{Word2vec}. This process is repeated a set number of times to generate a sequence of nodes, resembling a sentence in NLP. Random walks capture both local structure (directly connected nodes) and broader patterns (nodes connected through multiple steps), making them efficient for handling large graphs.

In embedding models for graphs, positive and negative edges play a crucial role in training. Positive edges represent actual, observed connections between nodes, while negative edges signify pairs of nodes without connections. Negative sampling is important because it balances the dataset, preventing the model from overemphasizing positive connections \citep{xu_understanding_2021}. Since graphs usually contain far more possible node pairs than actual edges, random sampling of negative edges is used to reduce computational demands. This balance in edge types enables the model to learn meaningful distinctions between connected and unconnected nodes, which enhances its predictive power.

The embeddings produced by these models are vectors that encapsulate the relationships between nodes based on their connections. These vectors can then serve as input to other machine learning models—such as a multilayer perceptron (MLP; e.g., \citealt{Rumelhart1986}) or other traditional models—enabling predictions or classifications. The process allows for applying neural network-based models to graph data, leveraging learned structural patterns and providing insights into the relational data embedded in the graph.

\section{Illustration}\label{Aplicación}

In this section, we analyze collaboration networks from Arxiv by comparing three link prediction models: ERGM, GCN, and Word2Vec. Our findings indicate that ERGM faces limitations with large networks, whereas the GCN model is the fastest, and the Word2Vec model offers the highest accuracy. These results provide empirical evidence that deep learning models are more effective for handling complex networks.

\subsection{Data}

To illustrate the methodologies presented in the previous section, we examine five collaboration networks representing scientific partnerships between authors of articles submitted to corresponding categories on the Arxiv platform:
\begin{itemize}
    \item \textbf{Astro-Ph}: Astrophysics, with 198,110 edges and 18,772 nodes.
    \item \textbf{Cond-Mat}: Condensed Matter, with 93,497 edges and 23,113 nodes.
    \item \textbf{Gr-Qc}: General Relativity, with 14,496 edges and 5,242 nodes.
    \item \textbf{Hep-Ph}: High-Energy Physics, with 118,521 edges and 12,008 nodes.
    \item \textbf{Hep-Th}: Theoretical High-Energy Physics, with 25,998 edges and 9,877 nodes.
\end{itemize}

These networks are undirected and unweighted (binary), indicating only whether two authors collaborated, without specifying the strength of collaboration. Since author characteristics are not included in the dataset, the analysis focuses primarily on relational data. This article centers its analysis on the Astro-Ph network, as it has the most connections. However, results are generally consistent across the other networks, as demonstrated later.

\subsection{Exploratory Analysis of the Astro-Ph Network}

In this section, we examine the structure of the Astro-Ph network. The network contains a few authors with a high number of connections, acting as central hubs. Naturally, this characteristic is uncommon: Only 59 individuals have more than 400 connections, representing just 0.31\% of all nodes. The average number of connections per author is approximately 18, and the maximum separation between two authors is 14 edges. The network's density of only 0.0022 indicates limited connectivity, which, in turn, suggests the presence of numerous cliques. The largest clique consists of 57 members, indicating the formation of distinct author groups, likely due to researchers’ preference for collaborating with colleagues from the same institution rather than with external partners. Finally, we observe that the network has 290 components, with the giant component including 95.37\% of the individuals.

\begin{figure}[htb]
    \centering
    \includegraphics[width = 0.6\textwidth]{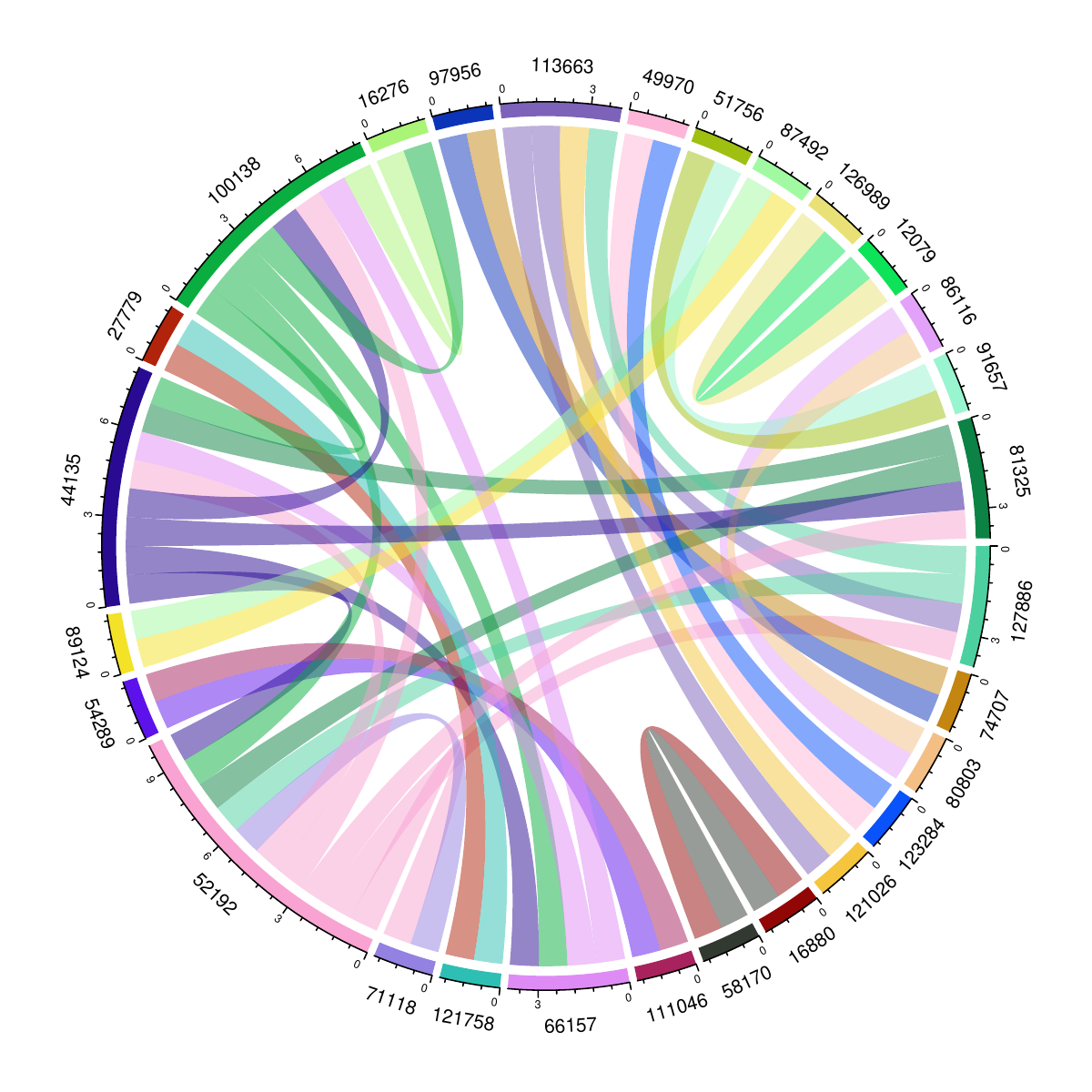}
    \caption{Chord diagram of a subset of authors in the Astro-Ph network.}
    \label{DiagramadeArco}
\end{figure}

Given the network’s size, visualizing the complete graph is impractical, so we display a subgraph of the 50 individuals with the highest degree, as shown in Figure \ref{DiagramadeArco}. First, we observe that a large portion of authors tends to collaborate with only a single co-author. Thus, researchers who work with multiple partners often choose collaborators within their immediate circle. This supports the conclusion that the collaborative network favors the formation of cliques. While many researchers collaborate with only one partner, those with more co-authors tend to work in more closed groups.

\subsection{Results}

First, the ERGM model is fitted to the large Astro-Ph network using the most basic configuration, which considers only links between nodes. This simple approach enables the model to handle a large network, but at the expense of performance, as it does not account for other important network features, such as triangles or other complex structures. When more structural features are included in the model formulation, the fitting algorithms fail to converge, even with distributed parallel computing over 20 cores, highlighting a clear limitation of this type of model.

The ERGM coefficient for the existence of links between nodes in the full Astro-Ph network is -6.7895, which is highly statistically significant. In natural scale, this coefficient indicates a 0.4998 probability of interaction between two authors. For the restricted network, this coefficient is positive and significant, suggesting a high likelihood of node connections while holding other factors constant. In this case, we also consider the coefficient associated with triangle formation, estimated at 9.9544, which indicates a high likelihood of links completing triads. Similarly, the coefficient for four-node star configurations is both positive and significant.

These findings underscore one of the key strengths of ERGMs: their capacity for providing interpretable results. However, given the dimensions of the networks that we consider here, ERGM execution times are exceedingly high. For the largest network, approximately nine hours are needed to fit the model and make link predictions. Although ERGMs generally offer a good fit, their low computational efficiency reflects that they are not well-suited for handling high-dimensional data.

To train the two remaining models, we define a balanced set of positive and negative edges, as the number of negative edges is much higher due to the network's low density. The second model implemented is a Graph Convolutional Network (GCN) model with two convolution layers. The first layer projects the input node features into a 64-dimensional latent space, applying the ReLU activation function to introduce non-linearity. The second layer also projects these representations into a 64-dimensional space, and finally, the output layer produces the model’s predictions, optimized using the Adam optimizer with a learning rate of 0.01. The binary link prediction loss is calculated using a combination of logarithmic losses for positive and negative edges. The model input consists of the edge indices, allowing the convolution to operate over the graph structure and capture topological information in the learned representations.

\begin{figure}[htb]
    \centering
    \includegraphics[width = 0.8\textwidth]{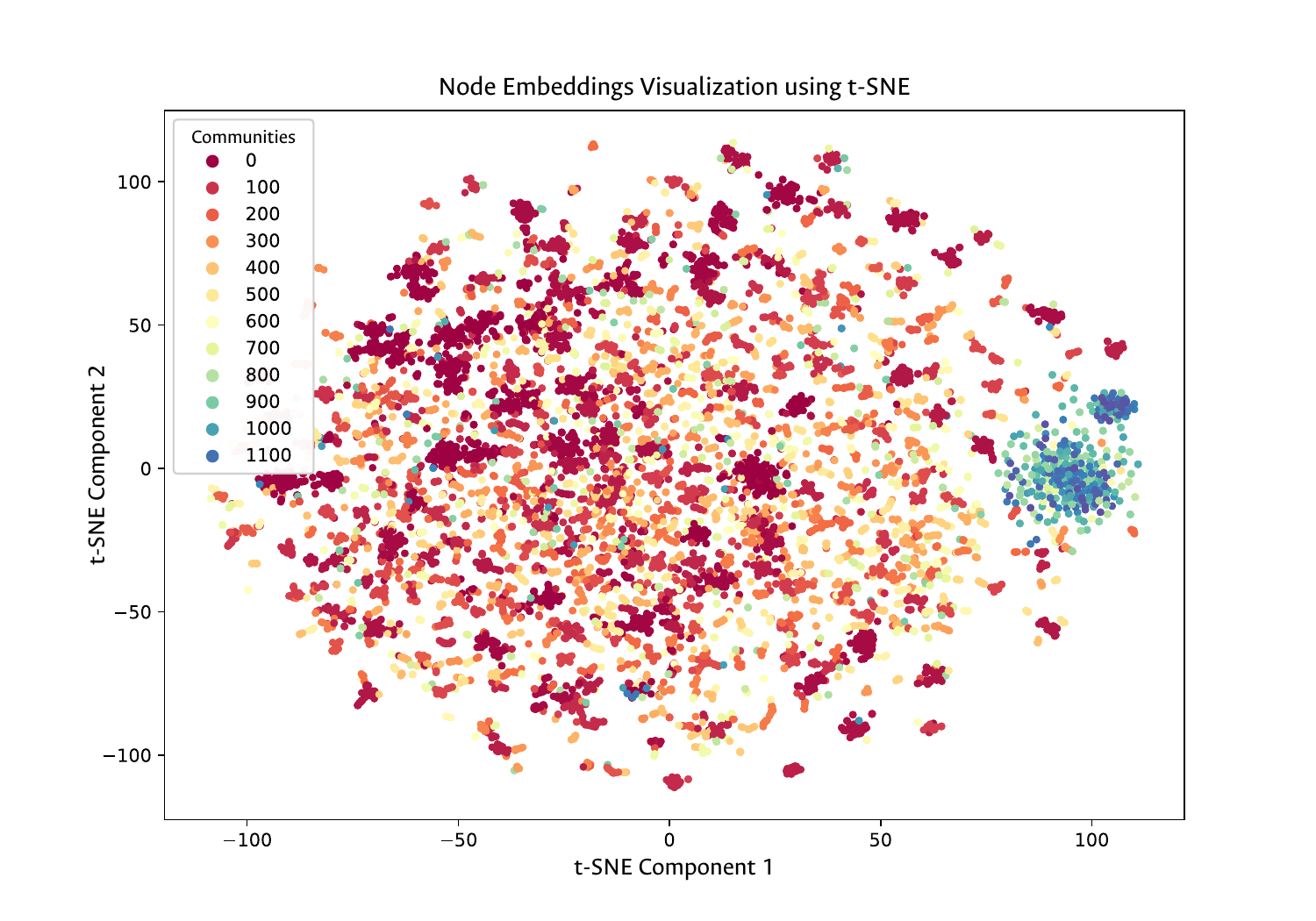}
    \caption{Visualization of embeddings and their corresponding clusters (using infomap) using t-SNE for the Astro-Ph network.}
    \label{repre}
\end{figure}

Finally, we fit the Word2Vec model using a DeepWalk-based approach to generate 100 random walks on the graph, each with a length of 30 nodes. The model is configured with a vector size of 32 dimensions, a contextual window of 10 nodes, a minimum count of 1, and a skip-gram approach, utilizing 4 workers to parallelize the process. The resulting node embeddings are then used as input for a multilayer perceptron (MLP). Figure \ref{repre} visualizes the Word2Vec model’s results by projecting the original 12-dimensional embeddings into a two-dimensional space using t-distributed Stochastic Neighbor Embedding (t-SNE; e.g., \citealt{vanDerMaaten2008}). In this projection, nodes from the same community in the original graph tend to cluster together, indicating that the embeddings effectively capture community structure despite the reduction from the original 12 dimensions.

Continuing with this process, the MLP model takes the embeddings generated by Word2Vec as input. This MLP has a two-layer architecture: The first layer contains 64 units and uses the ReLU activation function to introduce non-linearity, while the second layer projects to the appropriate output dimension. The model is optimized using the Adam optimizer with a learning rate of 0.01. As before, the binary link prediction loss is calculated using a combination of logarithmic losses for positive and negative edges.

To compare the models' performance, we use the area under the receiver operating characteristic (ROC) curve (AUC; e.g., \citealt{fawcett2006}) as a metric to quantify each model's ability to distinguish between classes, with values closer to 1 indicating higher performance. Additionally, we use a confusion matrix (e.g., \citealt{sokolova2009}) to observe the model's correct and incorrect predictions, categorized as true positives, false positives, true negatives, and false negatives, providing a detailed understanding of classification performance.

\begin{figure}[!htb]
    \centering
\includegraphics[scale = 0.38]{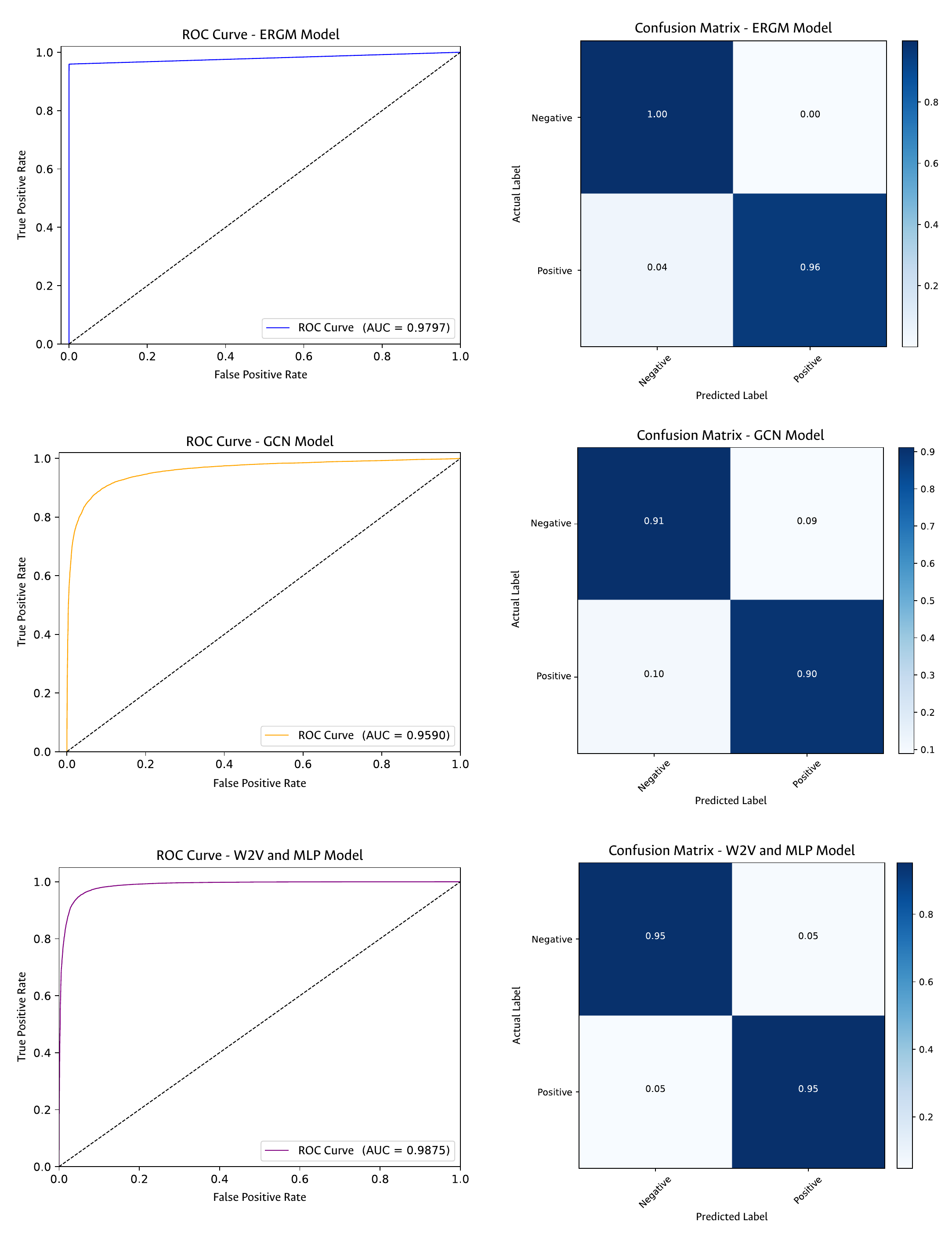}
    \caption{ROC curve and confusion matrix for the ERGM (first row), GCN model (second row), and Word2Vec model (third row) models.}
    \label{Comp}
\end{figure}

The results for the models using the Astro-Ph network are shown in Figure \ref{Comp}. For the ERGM, the AUC is 0.9797, indicating a high level of performance. The confusion matrix also shows that 96\% of positive class samples were correctly classified, while 4\% were incorrectly classified as negatives. Furthermore, this model had no false positives, which is advantageous for classification, and 100\% of negative class samples were correctly classified. The high accuracy in negative class classification (100\%), along with a high AUC and a significant percentage of false positives, suggests that the model generalizes well despite its substantial computational cost.

On the other hand, the GCN model achieves an AUC of 0.9590, which is quite good and competitive compared to the previous model. The confusion matrix reveals that 91\% of negative class samples were correctly classified, while 9\% were incorrectly classified as positives. These results indicate that the model performs well for negative edges (edges that do not exist in the network). For the positive class, 90\% were correctly classified, while 10\% were false negatives. These findings suggest that the model is effective in identifying positive edges (edges that do exist in the network). The AUC and confusion matrix show that the model performs well in distinguishing existing and non-existing connections within the network.

The Word2Vec model achieved an AUC of 0.9875, ranking it as the best-performing model among those evaluated. The confusion matrix shows that 95\% of negative class samples were correctly classified, while 5\% were incorrectly classified as positives. Additionally, for the positive class, 95\% were correctly classified, with 5\% classified as false negatives. These results demonstrate the model’s high effectiveness in identifying both positive and negative edges, with a low error rate in both cases. The AUC and confusion matrix confirm this model as the top performer in predictive accuracy.

Finally, when comparing models in terms of computational efficiency, measured by the time taken to fit the models to the datasets and generate predictions, the ERGM exhibits the poorest performance. For the Astro-Ph network, the ERGM takes over nine hours to execute, even with parallel computing methods, while the GCN model takes less than 8 seconds in total. Meanwhile, the Word2Vec model takes just over half an hour, placing it between the two. This highlights the capability of modern machine learning models to handle large volumes of data efficiently.

\begin{table}[htb]
    \centering
    \begin{tabular}{lcccccc}
        \toprule
         & \multicolumn{3}{c}{\textbf{AUC}} & \multicolumn{3}{c}{\textbf{Time}} \\
        \cmidrule(lr){2-4} \cmidrule(lr){5-7}
        & \textbf{ERGM} & \textbf{GCN} & \textbf{Word2Vec} & \textbf{ERGM} & \textbf{GCN} & \textbf{Word2Vec} \\
        \midrule
        Astro-Ph & 0.98 & 0.96 & 0.98 & 31,284.0 & 7.9  & 2,174.1  \\
        Cond-Mat & 0.96 & 0.91 & 0.99 & 18,360.0 & 4.7  & 2,594.4  \\
        Gr-Qc    & 0.78 & 0.89 & 0.99 & 3,307.8  & 1.1  & 472.2    \\
        Hep-Ph   & 0.97 & 0.95 & 0.99 & 11,844.0 & 2.9  & 1,257.7  \\
        Hep-Th   & 0.86 & 0.84 & 0.99 & 638.4    & 1.1  & 943.1    \\
        \bottomrule
    \end{tabular}
    \caption{AUC results and time (in seconds) for different networks and models.}
    \label{tab:auc_time}
\end{table}

The AUC and time taken for each network and model are reported in Table \ref{tab:auc_time}. The results show that, in terms of accuracy (AUC), the Word2Vec model achieves the highest scores, reaching 0.99 in most networks. Meanwhile, GCN also achieves high AUC values, particularly in the Hep-Ph and Gr-Qc networks, where its performance is similar to that of the Word2Vec model. In contrast, ERGM shows greater variability in the AUC values, with high values in some networks (such as Astro-Ph and Hep-Ph) and lower values in others (such as Gr-Qc). Regarding execution time, GCN is the fastest model, with significantly lower times compared to ERGM and Word2Vec, which are much slower, especially in larger networks like Astro-Ph and Cond-Mat. Overall, Word2Vec stands out in terms of accuracy, while GCN is notably more efficient in execution time. The codes used to fit the models can be found at \url{https://github.com/damartinezsi/An-unified-approach-to-link-prediction-in-collaboration-networks}.

\section{Discussion}

The high computational cost associated with fitting more complex ERGMs on large networks poses a considerable challenge. The inclusion of additional terms and more complex structures, such as triangles and other dependency patterns, increases the difficulty of fitting the model, often preventing it from meeting convergence criteria. In this context, taking a relatively small sub-sample of nodes is not a viable alternative, as it results in the loss of connection patterns present in the complete network, leading to unreliable predictions. This situation highlights the limited capacity of ERGMs to handle large-scale networks and underscores the need to develop and apply alternative strategies for addressing large networks without compromising model fit quality.

Machine learning models demonstrated outstanding performance in link prediction, underscoring their ability to capture complex patterns in network data. Unlike traditional approaches such as ERGM, which rely on statistical assumptions and are more suitable for smaller networks, deep learning models like GCN and Word2Vec models are designed to efficiently scale with large data volumes. This is due to their capacity to process and learn from both global and local graph structures through deep layers and vector embeddings, making them powerful and flexible tools for analyzing large-scale, high-dimensional networks.

Future work could involve comparing the implemented models with latent space-based approaches, which may provide an alternative for link prediction. Additionally, applying these models to networks with nodal attributes would be a promising direction, as it would allow for a more precise capture of the heterogeneity and dynamism of complex networks. This expansion could not only improve the predictive capability of the models but also provide a deeper understanding of the underlying interactions and structures in data-rich networks.

\section*{Statements and Declarations}

The authors declare that they have no known competing financial interests or personal relationships that could have appeared to influence the work reported in this article.

\bibliography{references.bib}
\bibliographystyle{apalike}


\end{document}